\documentclass{revtex4}
\usepackage{amsmath,amssymb,pstricks,epsfig}
\begin{document}
\setcounter{figure}{0}
\title{\bf Quasiparticle model for deconfined matter and the nucleation of hadrons}
\author{Andr\'e Bessa\footnote{abessa@if.ufrj.br} and Eduardo S. Fraga\footnote{fraga@if.ufrj.br}}
\address{Instituto de F\'\i sica, Universidade Federal do Rio de Janeiro\\
C.P.~68528, Rio de Janeiro, RJ 21941-972, Brasil}
\date{\today}


\begin{abstract}

Assuming a first-order chiral transition scenario, we study the process of phase conversion driven by homogeneous nucleation. We adopt a quasiparticle model whose parameters are fit to lattice QCD data to describe the pressure at high temperature in the deconfined sector, and a bag model equation of state for pions in the low-temperature sector. We compute the critical radius and nucleation rate in the thin-wall approximation, and compare the results to the ones obtained using the bag model and the linear $\sigma$ model. 
\end{abstract}

\maketitle


\strut It is widely accepted that QCD exhibits a transition from a confined hadronic phase to a deconfined partonic phase, the quark-gluon plasma (QGP). This plasma can presumably be produced under extreme conditions, such as in high-energy heavy-ion collisions. Furthermore, it is believed that the QGP existed in the early universe, a few seconds after the big-bang.

The character of the deconfining transition remains as an open question. Random matrix and effective model calculations suggest that, at low $\mu$  and non-zero $T$, a smooth crossover transition is expected and that, at low $T$ and non-zero $\mu$ , the chiral phase transition should be of first order. Thus, a second-order critical point must exist between these two limits.

Motivated by a recent analysis of elliptic flow\cite{pasi}, we assume a scenario of a first-order QCD transition\cite{pasi}. For simplicity, we work with two light quarks at $\mu=0$, with a critical temperature of 170MeV. The main idea is the use of an equation of state (EoS) for the high-temperature phase which is substantially better than the ideal gas one. The improved EoS is provided by the quasiparticle model\cite{peshierpavlenko,peshiersoff,schneiderweise}, and successfully reproduces lattice data for temperatures down to $T \approx T_{c}$. To describe the low-temperature phase we adopt a gas of massive pions.

For a small supercooling first order transitions are driven by the process of bubble nucleation\cite{csernaikapusta}. This is typical of slowly expanding systems as the early universe\cite{schwarz}. That hypothesis does not seem to apply to heavy-ion collisions, whose time scales are shorter than those in primordial universe by a factor of $\approx 10^{19}$. 

\normalsize

In this work we use the quasiparticle EoS to calculate the free energy, critical radius and nucleation rate. In this first approach we resort to the construction of a coarse-grained free energy and expressions valid within the thin-wall approximation. Finally, we compare our findings to bag model\cite{csernaikapusta} and chiral model results\cite{fragaadrian}. 

At temperatures around $T_{c}$ the coupling strength is large and perturbative calculations are inconsistent due to a strongly oscillating behavior\cite{zhai}. This is in conflict with lattice simulations where pressure varies smoothly down to low  temperatures. Such illness of the analytic calculation suggests one should look for more suitable lower-order approximations involving effective degrees of freedom (the quasiparticles) in the strongly interacting regime. 

The model assumes an ideal gas EoS of quark and gluon quasiparticles\cite{peshiersoff}
\begin{align}
p \,=\,\sum_{i} p_{id}^{(i)}\big (T, m_{i}^{2}(T) \big ) \; - \; B\big (m_{1}^{2}(T),m_{2}^{2}(T),\ldots \big) \; ,  
\end{align}
with thermally generated masses given by perturbative calculations using the hard thermal loop scheme\cite{lebellac},
\begin{align}
m_{g}^{2}(T) = \frac{1}{6}\bigg( N_{c}+ \frac{N_{f}}{2} \bigg )\, g^{2}T^{2} \hspace{1cm}\hbox{and}\hspace{1cm}m_{q} = \frac{N_{c}^{2}-1}{8N_{c}}\,g^{2}T^{2}\;,
\end{align}
where g is the leading-order QCD running coupling,
\begin{align}
g^{2}(T) = \frac{48 \,\pi^{2}}{(11N_{c}- 2N_{f})\,\log \big[(\lambda(T-T_{s})/T_{c}\big]^{2}}\;\;\; .
\end{align}

The parameter $T_{c}/\lambda$ is related to the QCD scale $\Lambda$, and $T_{s}$ accounts for the behavior in the infrared\cite{peshiersoff}. The function $B$ is necessary in order to preserve thermodynamical consistency. This is imposed through the condition\cite{peshiersoff} 
\begin{align}
\frac{\partial B}{\partial m_{j}^{2}} \;=\; \frac{\partial P_{id}}{\partial m_{j}^{2}}\;\;.
\end{align}

The pressure in the quasiparticle model reproduces with great precision lattice data even at temperatures near $T_{c}$, where the model was not expected to behave well. In this range the coupling is very large and the quasiparticles are much heavier.

\begin{figure}[t]
\centering
\includegraphics[width=8cm]{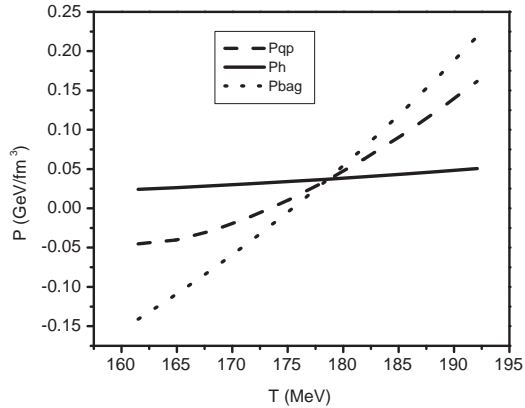}
\caption{Pressure curves around $T_{c}= 178$ MeV. The hadronic pressure is the solid line. The dashed line is from the quasiparticle model and the dotted one corresponds to the bag model.}\label{pqpandpbag}
\end{figure}

\begin{figure}[b]
\centering
\includegraphics[width=8cm]{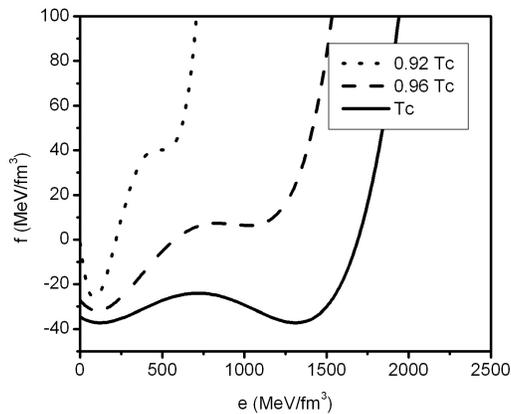}
\caption{The coarse-grained free energy at different temperatures ($T_{c}=178$MeV).}\label{energialivre}
\end{figure}

One needs to match the EoS for the plasma onto that for hadrons to study the transition. If we simply plot the quasiparticle EoS (adjusted with the lattice) together with that for a pion gas, there is no crossing at all. Bearing the comparison with the bag model in mind we argue that, although the quasiparticle EoS agrees with lattice predictions down to $T_{c}$, it does not represent the plasma phase immediately around $T_{c}$. The fast increase of the quasiparticle masses in this region is already an indicative of the phase transition, and in our model that information is carried by the low-temperature EoS.

Therefore, we avoid the region around $T_{c}$ by the introduction of a bag constant (set equal to $156$ MeV). This procedure makes the pressure curves cross at $T= 178$ MeV (see Fig. \ref{pqpandpbag}), still inside the physical range\cite{laermann}, without spoiling the data fit for higher temperatures.

In a coarse-grained theory the free energy has its minimum at the hadronic phase, for $T < T_{c}$,  and at the plasma phase, for $T > T_{c}$. We follow the usual method to interpolate between these extrema and to construct a free energy as a functional of the energy density\cite{csernaikapusta}. 
In the thin-wall approximation the height of the barrier is determined in terms of the surface tension and the correlation length of the static profile. The plot of Fig. \ref{energialivre} was obtained using typical values\cite{csernaikapusta,kajantie,huang}, namely $\sigma = 50$ MeV/fm$^2$ and $\xi = 0.7$ fm (ignoring any dependence on $T$).    

\begin{figure}[t]
\centering
\includegraphics[width=8cm]{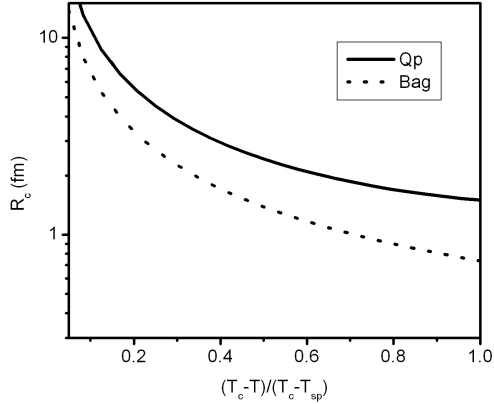}
\caption{Comparison of the quasiparticle and the bag model critical radii in the thin-wall approximation ($T_{c}=178$ MeV and $T_{s}=164$ MeV).}\label{rcritqpandbag}
\end{figure}

By inspection of the shape of the free energy we estimate the spinodal temperature, $T_{s}$, to be about $0.92 T_{c}$, that is $164$ MeV. Fig. \ref{rcritqpandbag} displays the critical radius, $R_{c}$, in the thin-wall approximation as a function of the supercooling. One can see that the critical radius for the quasiparticle EoS is always greater than the one for the bag model. This is a consequence of the smaller differences of the plasma and hadron pressures in the quasiparticle model as compared to the bag model (see Fig \ref{pqpandpbag}).
The nucleation rate is proportional to $\exp\{-\Delta F/T\}$, where $\Delta F$ is the energy cost for the formation of a critical bubble. Since $\Delta F \propto R_{c}^2$, the nucleation rate will be much smaller for the quasiparticle EoS than for the bag model. 

\begin{figure}[b]
\centering
\includegraphics[width=8cm]{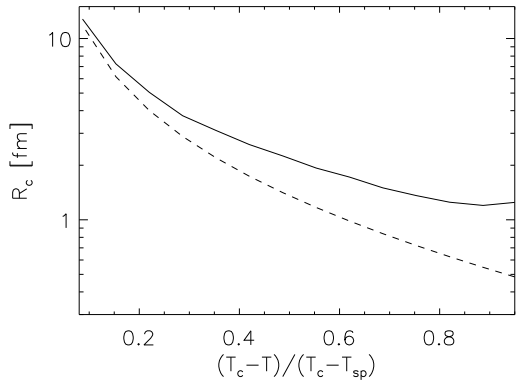}
\caption{The dashed line is the critical radius calculated with the linear $\sigma$ model coupled to quarks in the thin-wall approximation (extracted from \cite{fragaadrian}).}\label{rcritchiral}
\end{figure}

Fig. \ref{rcritchiral} shows the critical radius as a function of the supercooling obtained using the linear $\sigma$ model coupled to quarks\cite{fragaadrian}, with $T_{c}=123.7$ MeV and $T_{s}=108$ MeV. We observe that the chiral critical radius is also smaller than our quasiparticle result in the thin-wall approximation.

We conclude that the nucleation process of the QCD phase transition using the phenomenological quasiparticle EoS for the deconfined phase is qualitatively different from previous calculations using the bag model or approaches via chiral models. Our results are consistent with a {\it weakly} first-order phase transition, which seems to be a physically reasonable scenario\cite{pasi,laermann}. 
The use of the thin-wall approximation was a simplification in order to deal with analytic expressions, but the same calculation can be done numerically  under weaker assumptions. 

 We thank C.A.A. de Carvalho, A. Peshier and R. D. Pisarski for discussions. A.B. acknowledges CNPq for a PhD fellowship and the Millennium Institute for Quantum Information for financial support.  E.S.F is partially supported by CAPES, CNPq, FAPERJ and FUJB/UFRJ.

\end{document}